\documentstyle[12pt,epsfig,amsfonts,latexsym]{article}

\makeatletter

\@addtoreset{equation}{section} \makeatother

\newcommand{\HH}{{\cal H}}

\newcommand{\be}{\begin{equation}}
\newcommand{\ee}{\end{equation}}
\newcommand{\ba}{\begin{eqnarray}}
\newcommand{\ea}{\end{eqnarray}}
\newcommand{\bea}{\begin{eqnarray}}
\newcommand{\eea}{\end{eqnarray}}
\newcommand{\bean}{\begin{eqnarray*}}
\newcommand{\eean}{\end{eqnarray*}}
\newcommand{\nn}{\nonumber}

\newcommand{\bk}{{\bf k}}
\newcommand{\bx}{{\bf x}}

\def\spose#1{\hbox to 0pt{#1\hss}}
\def\lsim{\mathrel{\spose{\lower 3pt\hbox{$\mathchar"218$}}
\raise 2.0pt\hbox{$\mathchar"13C$}}}
\def\gsim{\mathrel{\spose{\lower 3pt\hbox{$\mathchar"218$}}
\raise 2.0pt\hbox{$\mathchar"13E$}}}

\begin{document}



\begin{flushright}
LPT-ORSAY 03-53
\end{flushright}
\vskip 1cm
\begin{center}
{\Large \bf Tachyon inflation: tests and comparison with single
scalar field inflation}
\end{center}
\vskip 1cm
\begin{center}
D.A.Steer $^{a}$\footnote{{\tt steer@th.u-psud.fr}} and F.Vernizzi
$^b$\footnote{{\tt vernizzi@iap.fr}}
\\
\vskip 5pt \vskip 3pt
{\it a}) Laboratoire
de Physique
Th\'eorique\footnote{Unit\'e Mixte de Recherche du CNRS (UMR 8627).}, B\^at. 210, Universit\'e
Paris XI, \\ 91405 Orsay Cedex, France\\
and
\\
F\'ed\'eration de recherche APC, Universit\'e Paris VII,\\
2 place Jussieu - 75251 Paris Cedex 05, France.
\vskip 3pt
{\it b}) Institut d'Astrophysique de
Paris, GR$\varepsilon$CO, FRE 2435-CNRS, 98bis boulevard Arago,
75014 Paris, France
\end{center}
\vskip 2.0cm

\setcounter{footnote}{0} \typeout{--- Main Text Start ---}



\begin{abstract}
We compare the standard single scalar field inflationary
predictions with those of an inflationary phase driven by a
tachyon field.  A slow-roll formalism is defined for tachyon
inflation, and we derive the spectra of scalar and tensor
perturbations as well as the consistency relations. At lowest
order the predictions of standard and tachyon inflation are shown
to be the same.  Higher order deviations are present and their
observational relevance is discussed. We then study some typical
inflationary tachyon potentials, discuss their observational
consequences and compare them with recent data. All the models
predict a negative and very small running of the scalar spectral
index, and they consistently lie within the 1$\sigma$ contour of
the data set.  However, the regime of blue scalar spectral index
and large gravity waves cannot be explored by these models.
Finally, a new exact solution of the unperturbed and perturbed
coupled gravity and tachyon equations is also presented.

\end{abstract}


\date{\today}

\renewcommand{\thefootnote}{\arabic{footnote}} \setcounter{footnote}{0}

\newpage

\section{Introduction}

The recent WMAP data
\cite{WMAP1,WMAP2,Peiris,WMAP4,WMAP5} strongly supports the idea
that the early universe underwent a phase of accelerated
expansion or inflation \cite{Guth}.
Inflation is becoming the dominant paradigm for the
generation of super-horizon fluctuations with a scale-invariant
spectrum, which are thought to be the origin of the large scale
structures.

One typically considers an inflationary phase driven by the
potential or vacuum energy of a scalar field, the inflaton, whose
dynamics is determined by the Klein-Gordon action
\cite{LiddleLyth}. More recently, however, motivated by string
theory, other non-standard scalar field actions have been used in
cosmology. In $k$-inflation \cite{Kinflation} higher-order scalar
kinetic terms in the action can, without the help of the
potential, drive an inflationary evolution. In this context,
models of quintessence such as $k$-essence may also resolve the
coincidence problem \cite{chiba,Kessence} (see, however,
\cite{Ed}). One particular model of $k$-inflation which has
recently attracted a great deal of attention is tachyon inflation
(see e.g.\ \cite{Gibbons1}), where the tachyon action is given by
\be
S_T = - \int d^4 x \sqrt{-g}  V(T)  \left( 1 +
g^{\mu \nu} \partial_{\mu} T
\partial_{\nu} T  \right)^{1/2} \label{eq:action1}
\ee
and the metric has signature $-,+,+,+$. The
tachyon $T$ is a real scalar field with dimensions of length and
$V(T)$ is its potential.

The motivations for studying the action Eq.~(\ref{eq:action1})
come from type II string theory. There the tachyon signals the
instability of unstable and uncharged D-branes of tension
$\lambda$, and different approaches \cite{action} have led to the
effective tachyon action being of the Dirac-Born-Infeld form given
in Eq.~(\ref{eq:action1}). In this context the positive potential
$V(T)$ is even, has a global maximum at $T=0$, and minima as $|T|
\rightarrow \infty$ where $V \rightarrow 0$. Different potentials
have been calculated, but one with particularly attractive
properties\footnote{Sen's conjecture \cite{sen2} is that the
static kink-like solutions of the tachyon action are the stable
D-brane into which the non-BPS brane decays. For the inverse cosh
potential these kinks have special properties
\cite{SolK,SolM,SolS}.} that will be studied in
Section~\ref{sec:models} is \cite{cosh,bsft,cosmocosh}
\be V(T) =
\frac{\lambda}{\cosh ({T}/{T_0})}. \label{eq:cosh}
\ee

Numerous papers (see for example
\cite{Gibbons1,Gibbons2,Malcolm,FKS,KL,Sami,Wasserman,Pad0}
and references within)
have investigated the cosmological
consequences of the gravity-tachyon system,
\be
S = \int d^4 x
\sqrt{-g} \frac{R}{16 \pi G} +S_T,
\label{eq:action}
\ee
including slow-roll inflation in the potential $V(T)$.  Indeed,
many potentials with the properties outlined above can drive
inflation, which typically takes place
at a scale characterized by the brane tension, $H \sim
\lambda^{1/2}/M_{\rm Pl}$, where $M_{\rm Pl} =(8 \pi G)^{-1/2}$.
Furthermore, Sen \cite{Sen3} has pointed
out that the rolling tachyon can contribute to the energy density
of the universe with dust-like equation of state, $P=0$. This has
raised the question of whether the tachyon could at the same time
drive inflation and later behave as dark matter.

As a possible mechanism for driving inflation, tachyon
condensation has been criticized in \cite{FKS,KL}. The main reason
is that for string theory motivated values of the parameters in $V(T)$,
there is an incompatibility between the slow-roll condition and
the COBE normalisation of fluctuations: Inflation generally takes place
at an energy scale
$\lambda^{1/4}$ with $T \sim T_0$, and in string theory $T_0 \sim 1/M_s$ where
$M_s$ is the string mass, and
\be
\lambda = \frac{M_s^4}{g_s (2
\pi)^3}
, \label{eq:string}
\ee
where $g_s$ is the string coupling.
The useful constant dimensionless ratio \cite{Malcolm}
\be
X_0^2 \equiv \frac{ \lambda T_0^2}{M_{\rm Pl}^2} \label{eq:X0}
\ee
appears in the slow-roll parameters derived from these potentials
and typically $X_0 \gg 1$ in order for the slow-roll conditions to
be satisfied. One can then see that for natural values of $g_s$
and $M_s$, slow-roll $X_0^2 \gg 1$ takes only place at an energy
scale which is too big to be compatible with the COBE constraint
$H/M_{\rm Pl} \sim \lambda^{1/2}/M_{\rm Pl}^2 \sim 10^{-5}$.
Potentials can, however, be found for which these issues may be
circumvented (see also e.g.~\cite{Bento,Mu} in the braneworld
context).

This criticism has cast a shadow on the string motivation of this
scenario but cannot deny the fact that a field satisfying the
action Eq.~(\ref{eq:action}) with $V(T)$ describing an
instability, can naturally lead to inflation. However, as in
standard inflation, one needs a small parameter (in this case
$(T_0 M_{\rm Pl})^{-1}$) in order to have a successful
inflationary phase.

Thus, despite this criticism and regardless of the string
motivations, here we take a phenomenological approach and study
the inflationary predictions of a phase of inflation driven by a
field $T$ satisfying the action Eq.~(\ref{eq:action}). We call
this tachyon inflation although the potential $V(T)$ may not be
particularly string inspired.  However, throughout this paper we
assume that $V(T)$ satisfies the properties mentioned above,
namely
\be
V(0)=\lambda, \qquad V'(T>0)<0, \qquad V(|T| \rightarrow \infty)
\rightarrow 0. \label{eq:propsV}
\ee
The questions we address here are: 1) Does tachyon inflation lead
to the same predictions as standard single field inflation (SSFI)?
2) Can tachyon inflation already be ruled out by current
observations? 3) Can we discriminate between tachyon inflation and
SSFI in the light of new and planned future experiments?

The answer to the first question is no: tachyon inflation leads to
a deviation in one of the second order consistency relations.
However, the answer to the second question is that tachyon
inflation cannot be ruled out at the moment, and its predictions
are typically characteristic of small field or chaotic inflation.
The answer to the final question may also be negative: no
characteristic signatures of tachyon inflation are likely to be
detectable by planned observations but this may change in the
future.

Before concluding this section, a comment is in order here. As
opposed to action (\ref{eq:action1}), the linear action
\be
S_T = - \int d^4 x \sqrt{-g}  V(T)  \left( 1 +
g^{\mu \nu} \partial_{\mu} T
\partial_{\nu} T  \right) \label{eq:actionSSFI}
\ee
can be put into the standard Klein-Gordon form for a scalar field
$\phi$: let
\be
T=T(\phi) \qquad {\rm with} \qquad d \phi = \sqrt{2V(T)} \; dT.
\label{eq:change1}
\ee
Then the corresponding potential for $\phi$ is
\be
W(\phi)=V(T(\phi)) \label{eq:change2}
\ee
so that the inflationary predictions of action
(\ref{eq:actionSSFI}) are the same as those of SSFI (see the
Appendix, Sec.~B). When the square root is present a similar
change of variables cannot be found, though the square root can be
linearized at the expense of introducing an auxiliary field which
can either be another scalar field \cite{Ed1} or a metric field
\cite{SolM}. Here we compare tachyon inflation with SSFI, and
hence any expansion of action (\ref{eq:action1}) in powers of
$\partial_\mu T \partial^\mu T$ must go beyond the first order
term in order for differences to be found.

This paper is set up in the following way.  In
Section~\ref{sec:unpert} we consider the unperturbed tachyon
system coupled to gravity.  We present (Section~\ref{sec:exact}) a
new exact solution to Eq.~(\ref{eq:action}) which shows explicitly
the inflationary and dust-like properties of the solution.
Slow-roll parameters are derived in Section~\ref{sec:inf}, where
we use the definition introduced in \cite{Dominique}.  This is the
natural one when comparing models in which inflation is driven by
different types of fields.  The spectra of scalar and tensor
perturbations and the running of the spectral indexes are also
derived in Section~\ref{sec:inf}. We show that one of the next to
lowest order consistency relations is different from the one
predicted by SSFI. In Section~\ref{sec:models}, different
potentials $V(T)$ are studied and their predictions compared with
recent data. In the Appendix, Sec.~B, we review and clarify the
large and small scale perturbations of a tachyon fluid.  In fact,
inorder to highlight the differences and similarities between
tachyon inflation and SSFI, in the Appendix, Sec.~A, we consider
the slightly more general action $ S = - \int d^4 x \sqrt{-g} V(T)
\left( 1 + g^{\mu \nu} \partial_{\mu} T
\partial_{\nu} T  \right)^{q}
$. In the following we often denote
\be
x \equiv T/T_0;
\ee
an overdot denotes a derivative with respect to cosmic time
$t$, and a prime a derivative with respect to the tachyon $T$.

\section{Unperturbed tachyon evolution} \label{sec:unpert}

We begin by reviewing the background properties of a tachyon
dominated universe. From (\ref{eq:action}), the energy momentum
tensor for the tachyon is given by
\be
T_{\mu \nu} = - V(T) g_{\mu \nu} \sqrt{1 + \partial^\alpha T
\partial_\alpha T} + \frac{V(T)}{\sqrt{1 + \partial^\alpha T
\partial_\alpha T}} \partial_\mu T
\partial_\nu T.
\label{eq:emttachyon}
\ee
Now split the tachyon field into a homogeneous time dependent
contribution, $T(t)$, and a small $\bx$-dependent
perturbation, $\delta T(t, \bx)$, which describes its quantum
fluctuations. In this section we summarize the basic equations for
$T(t)$.

The tachyon field can be
treated as a fluid with
\bea
T_{\mu \nu} &=& (\rho+P) u_{\mu} u_{\nu} + P g_{\mu \nu}, \\
u_\mu &=& \frac{\partial_\mu T}{ (-\partial_\alpha T \partial^\alpha T)^{1/2}},
\eea
where $\rho$, $P$ and $u_{\mu}$ are the density, pressure and
four-velocity of the fluid, respectively. In a homogeneous and isotropic
background, with line element $ds^2 = - dt^2 + a(t)^2 d \bx^2 $,
\bea
\rho  &=& \frac{V(T)}{(1 - \dot T^2)^{1/2}}, \label{eq:rho}\\
P &=& - V(T) (1 - \dot T^2)^{1/2} \label{eq:P}, \eea so that the
Friedmann equation is \be H^2 = \frac{1}{3M_{\rm Pl}^2}
\frac{V}{(1 - \dot T^2)^{1/2}}, \label{eq:frie} \ee whilst
energy-momentum conservation gives a second order equation for
$T(t)$, \be \frac{\ddot T}{1-\dot T^2} +  3 H \dot T +  ({\ln V})'
=0. \label{eq:kg} \ee For $\dot T \ll 1$ (i.e.~during inflation,
see below), Eq.~(\ref{eq:kg}) reduces to a Klein-Gordon equation
for a homogeneous scalar field with $\ln V$ playing the r\^ole of
the potential.

It will be convenient to define the ratio
\be w
\equiv \frac{P}{\rho}=-1+\dot T^2,
\ee
the adiabatic sound speed,
\be
c_A^2 \equiv \frac{\dot P}{\dot \rho} = -w \left(1 + \frac{2}{3}
\frac{(\ln V)'}{ H \dot T} \right), \label{eq:sstachyon} \ee
and the effective sound speed \cite{GarriMuka},
\be
c_S^2 \equiv \frac{\partial P/\partial \dot{T}^2}
{\partial \rho/\partial \dot{T}^2} = -w,
\ee
which takes into account dissipative effects\footnote{Equivalently
$c_S^2 = {\partial P}/{\partial \rho}$ with equation of state
$P = - V(T)^2/\rho$.}.  For adiabatic
perturbations the two sound speeds are the same but in
general they are different (see Appendix, Sec.~\ref{sec:pert}).

We conclude this section with a remark. For a number of
potentials,
such as the inverse cosh potential, at early times and small $T$,
$\dot{T}^2 \ll 1$ and $w=-1$, whereas at late times and large $T$,
$\dot{T}^2 \to 1$ and $w=0$.\footnote{For a large class of
potentials, including exponentially decaying potentials, this
dust-like behaviour is an attractor solution \cite{Wecht}.}
Furthermore, as summarized in the Appendix, as opposed to a
Klein-Gordon scalar field, $T$ can cluster on small scales. For
these reasons the tachyon has been studied as a potential unified
inflation--dark-matter candidate \cite{Gibbons1,Gibbons2,FKS,KL}.
This proposal, however, runs into some difficulties \cite{KL}: for
string-theory motivated potentials, it is not clear how to reheat
the universe after inflation (see, however, \cite{Cline:2002it}).
Note that in the brane world context reheating can occur very
efficiently through gravitational particle production
\cite{Bento,Ed3}. Furthermore, fine tuning is require for the
energy density in the tachyon to take the correct value today
\cite{Wasserman}.

\section{An inflation to dust exact solution}
\label{sec:exact}

Though it is somewhat outside the main theme of this paper, here
we present a new (and potentially useful) exact solution of the
coupled tachyon and gravity equations, Eqs.~(\ref{eq:frie}) and
(\ref{eq:kg}), which interpolates between the inflationary and
dust phases. Though other exact solutions have been presented in
the literature
\cite{Pad1,Feinstein}, their corresponding potentials do not
satisfy the properties given in Eq.~(\ref{eq:propsV}) and do not
probe both the inflationary and dust phases.

The form of Eq.~(\ref{eq:kg}) suggests the ansatz
\be
\dot{T}  = \tanh (t/T_0),
 \label{eq:tanhTdot}
\ee
where $T_0$ is a constant.  (Without loss of generality we
consider $T \geq 0$.)  Integrating gives
\be
\cosh (t/T_0) = e^{T/T_0},
 \label{eq:tT}
\ee
where we have taken $T(0)=0$.
From Eqs.~(\ref{eq:frie}) and (\ref{eq:kg}) it follows
that $V(t)$ satisfies
\be
\frac{\dot{V} \cosh (t/T_0)}{V \sinh (t/T_0)} +
\frac{\sqrt{3V}}{M_{\rm Pl}}  \frac{\sinh (t/T_0)}{\sqrt{\cosh
(t/T_0)}}  = - \frac{1}{T_0},
\ee
with solution
\be
V(t) = \frac{\lambda}{\cosh (t/T_0)} \left[\frac{1}{ 1 +
\frac{\sqrt{3}X_0}{2}  \left(t/T_0 -\tanh (t/T_0)\right)}
 \right]^2.
\label{eq:Vt}
\ee
Here we have normalized the potential such that $V(0)=\lambda$ and
$X_0$
is defined in (\ref{eq:X0}).

On using Eq.~(\ref{eq:tT}), $V(T)$ is thus of the form
\be
V(T) = \lambda e^{-T/T_0} f(T/T_0,X_0), \label{eq:blimey}
\ee
where $f(T)$ is a decreasing function of $T$.  For $T \ll T_0$
(when $t \simeq T_0 \sqrt{2T/T_0}$),
\be
f(T/T_0,X_0) \simeq 1 - \sqrt{\frac{8}{3}}X_0 \left(
\frac{T}{T_0} \right)^{3/2},  \ \ \ \ T/T_0 \to 0,  \label{eq:blimey2}
\ee
whereas for $T \gg T_0$ (when $t \sim T$),
\be
f(T/T_0,X_0) \simeq \frac{4}{3X_0^2} \left( \frac{T_0}{T}
\right)^2, \ \ \ \ T / T_0 \to \infty.
\ee
For small $T$, $ V \simeq \lambda \exp(-T/T_0)$, and the scale of
the potential is set by $\lambda$.  On the other hand, for large
$T$, $V \simeq ( M_{\rm Pl}^2/ T_0^2)  \exp(-T/T_0)\left( T_0/T
\right)^2$ and the scale is set by $M_{\rm Pl}^2/T_0^2$.
For
dimensional reasons, this non-perturbative dependence on $\lambda$
means that $V$ depends explicitly on $M_{\rm Pl}$.

Finally, combining Eqs.~(\ref{eq:frie}), (\ref{eq:tT}) and
(\ref{eq:Vt}) gives the Hubble parameter
\be
H^2 = \frac{X_0^2}{3 T_0^2} \left[\frac{1}{ 1 + \frac{\sqrt{3} X_0}
{2 } \left(t/ T_0 - \tanh (t/T_0)\right)}
 \right]^2,
 \label{eq:Hother}
\ee
which is unfortunately not analytically integrable for $a(t)$.
However, for large $T$, where the potential vanishes
exponentially and $\dot{T} \rightarrow 1$,
it follows from Eq.~(\ref{eq:Hother}) that $H =
2/3t$ and hence that $a \propto t^{2/3}$.  Thus again we find the
dust-like solution in this limit.  For small $T$ when
$\dot{T} \rightarrow 0$, $H
=\sqrt{\lambda}/\sqrt{3} M_{{\rm Pl}}$ and the universe inflates
exponentially.  Hence this solution interpolates between the two
behaviours of the tachyon fluid --- the inflationary and dust-like
ones.
In the remainder of this
paper we focus on the inflationary phase.

\section{Predictions of tachyon inflation}\label{sec:inf}

It is well known that during an inflationary phase, quantum vacuum
fluctuations are stretched on scales larger than the horizon.
There they are frozen until they reenter the horizon after
inflation. Regardless of the field which drives inflation, large
scale perturbations with a quasi scale invariant
(Harrison-Zel'dovich) spectrum are generated. Deviations from the
scale invariance can be measured in terms of the slow-roll
parameters which we define in Subsection~\ref{sub:B}. The spectra
of scalar and tensor perturbations generated during tachyon
inflation are expressed in terms of the slow-roll parameters in
Subsection~\ref{sub:D}. We show that at lowest order the spectrum
of scalar perturbations is the same as that of SSFI. Next to
lowest order corrections are also computed, and they change the
consistency relations with respect to those of SSFI
(Subsection~\ref{sub:F}).  In the next section these results will
be applied to particular models of tachyon inflation and compared
to recent data.

\subsection{Condition for tachyon inflation and slow-roll parameters}
\label{sub:B}

For tachyon inflation, the basic condition for accelerated
expansion is that
 \be \frac{\ddot a }{a} = -
\frac{1}{6 M_{\rm Pl}^2}(\rho + 3 P) =  \frac{1}{3 M_{\rm Pl}^2}
\frac{V}{(1-\dot{T}^2)^{1/2}} \left(1 -\frac{3}{2} \dot T^2
\right)
>0
\ee
so that $\dot T^2 < \frac{2}{3}$. From Eq.~(\ref{eq:kg}),
inflation will last sufficiently long if $\ddot T$ is smaller than
the friction term due to the expansion,
\be
\ddot T < 3 H\dot T. \label{eq:secondslowroll}
\ee
Thus as in SSFI, tachyon inflation is based upon the slow
evolution of $T$ in its potential $V(T)$, with the slow-roll
conditions
\be
\ddot T \ll  3H \dot T  , \ \ \ \    \dot T^2 \ll 1, \label{eq:src}
\ee
so that during inflation
\be
\dot{T} \sim - \frac{(\ln V)'}{3H}, \qquad H^2 \sim \frac{V}{3
M_{\rm Pl}^2}.
\ee

Now we want to define slow-roll parameters for tachyon inflation.
The slow-roll approximation is an expansion in terms of (generally
small) quantities derived from appropriate derivatives either of
the Hubble parameter $H$ or of the potential $V(T)$. There are
several ways in which to define the Hubble slow-roll parameters
for tachyon inflation. Here we use the recently introduced
horizon-flow parameters \cite{Dominique} based on
derivatives of $H$ with respect to the number of $e$-foldings $N$,
\be
N(t)  \equiv \int_t^{t_e} H(t) dt, \label{eq:defN}
\ee
where $t_e$ is the end of inflation. Note that for $\dot T>0$
\be
dT= -\frac{2}{3}  \frac{H'}{H^3} dN.
\label{eq:efoldT}
\ee
These slow-roll parameters are defined as
\bea
\epsilon_0 & \equiv & H_* / H ,\label{eq:defeps1} \\
\epsilon_{i+1} & \equiv & \frac{d \ln |\epsilon_i|}{dN}, \ \ \ i
\ge 0, \label{eq:defeps2}
\eea
where $H_*$ is the Hubble parameter at some chosen time, and
\be
\dot \epsilon_i = H \epsilon_i \epsilon_{i+1} . \label{eq:doteps}
\ee
Since these definitions are independent of the field driving
inflation, they are a natural choice to use in order to compare
SSFI and tachyon inflation.  They form exactly the same
hierarchy of inflationary flow equations as in SSFI (see indeed
\cite{Liddleflow}),
though the
observables (such as spectral indices) will no longer be related
to the $\epsilon_i$ in the same way.  In terms of $T$, the
definitions (\ref{eq:defeps1}) and (\ref{eq:defeps2}) are
\bea \epsilon_1   &=&  \frac{3}{2} \dot T^2
, \label{eq:eps} \\
\epsilon_2   &=& \sqrt{\frac{2}{3 \epsilon_1}} \frac{\epsilon_1'}{H}
=  2 \frac{\ddot T}{H \dot T} ,\\
\epsilon_2 \epsilon_3 &=& \sqrt{\frac{2 \epsilon_1}{3 }} \frac{\epsilon_2'}{H}.
\eea
(Note that $\sqrt{\epsilon_1}=-\sqrt{\frac{2}{3}}(H'/H^2)$ since
$H'=-\frac{3}{2}H^2 \dot T <0$.)
The first parameter $\epsilon_1$ measures the contribution of
$\dot T^2$ to the energy density Eq.~(\ref{eq:rho}) and, as in
SSFI, inflation occurs when $\epsilon_1 <1$, ending once
$\epsilon_1$ exceeds unity. The parameter $\epsilon_2$ measures
the ratio of the field acceleration relative to the friction
acting on it due to the expansion. The slow-roll conditions,
Eq.~(\ref{eq:src}), are satisfied when $\epsilon_1, \epsilon_2 \ll
1$.

The Friedmann equation and the conservation equation,
Eqs.~(\ref{eq:frie}) and (\ref{eq:kg}), can be rewritten as a
Hamilton-Jacobi system,
\bea
&& H'^2 - \frac{9}{4} H^4(T) +
\frac{1}{4} \frac{1}{M_{\rm Pl}^4}
V^2(T)=0, \label{eq:HJ2} \\
&&H' = -\frac{3}{2} H^2(T) \dot T, \label{eq:HJ1}
\eea
so that as in SSFI \cite{LPB}, the tachyon inflation solution is
an attractor. Note again that only up to first order terms in
$\epsilon_1$ is the inflationary dynamics
the same as in SSFI. This can be seen by comparing the Friedmann
equation for a single scalar field
\be
H^2(\phi) \left[1-\frac{1}{3} \epsilon_1 (\phi) \right] =
\frac{1}{3M_{\rm Pl}^2} W(\phi) \label{eq:friephi},
\ee
where $W(\phi)$ is the potential of $\phi$ (see Eq.~
(\ref{eq:change2}), with the one for the tachyon,
Eq.~(\ref{eq:frie}),
\be
H^2(T) \left[1-\frac{2}{3} \epsilon_1 (T)\right]^{1/2} = H^2(T)
\left[1-\frac{1}{3} \epsilon_1 (T) \right] + {\cal O}(\epsilon_1^2)=
\frac{1}{3M_{\rm Pl}^2} V(T) \label{eq:frieTT}.
\ee

On deriving Eq.~(\ref{eq:frieTT}) with respect to $T$, the tachyon
potential and its first and second derivatives can be expressed in
terms of the slow-roll parameters,
\bea
\frac{V'}{V H } &=& - {\sqrt{6 \epsilon_1}\left(1 -
\frac{2\epsilon_1}{3} + \frac{\epsilon_2}{6} \right)}
\left(1-\frac{2 \epsilon_1}{3} \right)^{-1}, \label{eq:VT/T}
\\
\frac{V''}{V H^2} &=&  3 \left(\epsilon_1 -
\frac{\epsilon_2}{2}\right) + \frac{\epsilon_2}{2} {\left(5
\epsilon_1 - \frac{\epsilon_2}{3} - \epsilon_3 \right) }{ \left(1 -
\frac{2}{3} \epsilon_1 \right)^{-1}}
\nn
\\
&&  + \; \; {6 \epsilon_1 \left( 1 - \frac{2}{3}\epsilon_1 +
\frac{\epsilon_2}{6} \right) \left( 1 - \frac{2}{3}\epsilon_1 -
\frac{\epsilon_2}{3}\right) }{\left(1 - \frac{2}{3}
\epsilon_1\right)^{-2}}. \label{eq:VTT/T}
\eea
At second order in the slow-roll parameters, these expressions are
different from those of SSFI. At leading order we have\footnote{
In SSFI with potential $W(\phi)$ related to $V(T)$ by
Eqs.~(\ref{eq:change1}) and (\ref{eq:change2}), one also finds the
following equations up to a multiplicative factor of 1/2 (see
Appendix, Sec.~B).}
\bea
\epsilon_1 &\simeq&
\frac{M_{\rm Pl}^2}{2}  \frac{V'^2}{V^3},  \label{eq:epsilonV}\\
\epsilon_2 &\simeq&  {M_{\rm Pl}^2} \left(-2
\frac{V''}{V^2}
+3  \frac{V'^2}{V^3}  \right), \label{eq:etaV} \\
\epsilon_2 \epsilon_3 &\simeq& {M_{\rm Pl}^4} \left(2
\frac{V''' V'}{V^4} -10
 \frac{V'' V'^2}{V^5} + 9 \frac{V'^4}{V^6}\right), \label{eq:xiV}
\eea
which should be compared with other definitions given in the
literature \cite{Malcolm}.

In terms of the slow-roll parameters, the number of $e$-foldings
is given by
\be
N(T)=  \sqrt{\frac{3}{2}} \int_{T}^{T_e}
\frac{H}{\sqrt{\epsilon_1}} dT  \simeq   \frac{1}{M_{\rm Pl}^2}
\int_{T_{e}}^{T} \frac{V^2}{V'} dT. \label{eq:NNN}
\ee
Both the slow-roll parameters and the spectra and amplitude of
perturbations are functions of $T$. Let $T_*$ denote the value of
$T$ at which a length scale crosses the Hubble radius during
inflation, $k=a_* H_*$. Then the definition of the number of
$e$-foldings, Eq.~(\ref{eq:defN}), gives
\be
a_* = a_e \exp(-N_*), \ \ \ \ k=a_* H_* = a_e H_* \exp(-N_*),
\ee
leading to
\be
\frac{d \ln k}{dT} =-\frac{3}{2} \frac{H^3}{H'} (1-\epsilon_1 ),
\label{eq:dk}
\ee
at $T=T_*$.
Finally, following \cite{Liddleetal}, the comoving scale $k$ can be
related to $N_*=N(T_*)$ through
\be
N_* = 62 - \ln \frac{k}{a_0 H_0} - \ln \frac{10^{16} {\rm
GeV}}{V(T_*)^{1/4}} + \ln \frac{V(T_*)^{1/4}}{V_e^{1/4}} - \gamma
\ln \frac{V_e^{1/4}}{\rho_{reh}^{1/4}}.
\ee
Here we assume that at the end of inflation when $V=V_e$ the
universe is reheated by some unknown mechanism, and the energy
density of the universe $\rho_{reh}$ is dominated by radiation.
Between the end of inflation and the end of the reheating, the
scale factor is assumed to evolve approximately as $a(t) \propto
t^\gamma$ with $0<\gamma<1$. A model independent upper
bound has been given for $N_*$ \cite{Hui}: it is
\be
N_* < 62.5 + \ln \left( \frac{0.6}{h} \right),
\ee
on observationally relevant scales, e.g.~when the scale $k=0.002 \
{\rm Mpc}^{-1}$ was crossing the horizon. More possibilities are
discussed in \cite{SamLiddle} where the authors find
\be
N_* = 63.3 + \frac{1}{4} \ln \epsilon_1,
\ee
for instantaneous reheating, but in extreme cases one can even
have $N_* \simeq 100$. Since after tachyon inflation the dynamics
of the reheating is still unclear, in the following we shall
assume a conservative value of $40 \le N_* \le 70$. Since a
further unknown parameter (i.e. $X_0$) is present in the analysis
of tachyon inflation, increasing the range of values of $N_*$ does
not change qualitatively our analysis.

\subsection{Calculating the density perturbations}
\label{sub:D}

Calculation of the spectra of scalar quantum fluctuations proceeds
by defining a canonical variable which can be quantized with the
standard methods. The straightforward generalization of the
canonical variable to the case of a tachyon fluid is (see
\cite{FKS,GarriMuka,Hwang}),
\be
{\rm v}_\bk=z M_{\rm Pl} {\cal R}_\bk,
\label{eq:defv}
\ee
where ${\cal R}_{\bk}$ is the curvature perturbation defined in the Appendix,
Eq.~(\ref{eq:curv}), and
where the pump field $z$ is defined by
\be
z = \frac{\sqrt{3} a \dot T}{(1 - \dot T^2)^{1/2}}
= \frac{a \sqrt{2 \epsilon_1}}{c_S}.
\label{eq:z}
\ee
The equation derived from minimizing the action expanded to second
order in $\rm v_\bk$ is \cite{GarriMuka}
\be
\frac{d^2{\rm v}_\bk}{d\tau^2} +\left(c_S k^2 - U(\tau) \right)
{\rm v}_\bk = 0, \ \ \ \ U(\tau) \equiv \frac{1}{z} \frac{d^2 z}{d
\tau^2}. \label{eq:canon}
\ee
It is important to note the factor of $c_S=-w$  in front of $k^2$.
Before
computing $U(\tau)$ in terms of the slow-roll parameters, we
observe that in SSFI inflation the pump field is
\cite{Mukhanovetal} (see also the Appendix),
\be
z_{\rm SSFI} \equiv a \sqrt{2 \epsilon_1}= z\left(1-\frac{2}{3}
\epsilon_1 \right)^{1/2}, \label{eq:zcomp}
\ee
differing from $z$ by a first order term in $\epsilon_1$.
It
follows that
\bea
U &=& U_{{\rm SSFI}} + \frac{2}{3} \left( \frac{1}{z} \frac{dz}{d
\tau} \frac{d \epsilon_1}{d \tau} + \frac{1}{2}\frac{d^2
\epsilon_1}{d \tau^2} \right) \left(1-\frac{2 }{3} \epsilon_1
\right)^{-1}
 + \frac{1}{9}
\left( \frac{d \epsilon_1}{d \tau} \right)^2
\left(1-\frac{2}{3} \epsilon_1 \right)^{-2} \nonumber \\
&=& U_{{\rm SSFI}} + a^2 H^2
\epsilon_1 \epsilon_2 + {\cal O} (\epsilon_i^3), \label{eq:zssfi}
\eea
where we have used Eq.~(\ref{eq:doteps}) and $dz/d\tau \simeq zaH$
at lowest order.
Therefore, up to first
order in $\epsilon_1$ and $\epsilon_2$, the term $U(\tau)$ in
Eq.~(\ref{eq:canon}) is the same as the one appearing in SSFI,
$U_{\rm SSFI}(\tau)$. The correction $ \propto \epsilon_1
\epsilon_2$ allows us to compute $U$ up to second order in the
slow-roll expansion from $U_{\rm SSFI}$ given in \cite{Dominique}.
Alternatively, $U(\tau)$ can be computed directly from the
expression, \ba U &=& 2 a^2 H^2 \left\{ 1 + \frac{3}{2} \dot T^2
\left(1+ \frac{V'}{H V \dot T} \right) + \frac{V'^2}{H^2 V^2} -
\frac{1}{2}\frac{V''}{H^2V} \right\} \label{eq:exactU}
\\
& = & a^2 H^2 \left(2 - \epsilon_1 +\frac{3}{2} \epsilon_2 + \frac{1}{4}
\epsilon_2^2 + \frac{1}{2} \epsilon_1 \epsilon_2
+ \frac{1}{2} \epsilon_2 \epsilon_3 \right) + {\cal O}(\epsilon_i^3).
\label{eq:expan}
\ea
Though $U$ given in Eq.~(\ref{eq:exactU}) can easily be calculated
exactly using Eqs.~(\ref{eq:VT/T}) and (\ref{eq:VTT/T}), we have
only written down explicitly the second order expression in (\ref{eq:expan}).
This is
sufficient to derive the spectrum of perturbations and the
consistency equations up to second order. As usual, the zeroth
order term $2a^2 H^2$ ensures that the spectrum is
scale invariant.

Following standard procedures, we look for a solution to
Eq.~(\ref{eq:canon}) by first expressing the conformal time parameter
$\tau$ as \cite{Dominique}
\be \tau = - \frac{1}{a H(1-\epsilon_1)} +
\int \frac{\epsilon_1 \epsilon_2 }{(1 - \epsilon_1)^2} \frac{dN}{a
H}= - \frac{1}{a H} (1+\epsilon_1) + {\cal
O}(\epsilon_i^2)\label{eq:tauofaH}.  \ee As given by
Eq.~(\ref{eq:doteps}), to first order $\epsilon_1$ is constant and
$aH\tau$ can be taken to be constant for each $k$ mode (though not
necessarily the same constant for different $k$).  On substituting
$aH$ from Eq.~(\ref{eq:tauofaH}) into the first order expansion
Eq.~(\ref{eq:expan}), Eq.~(\ref{eq:canon}) becomes \be
\frac{d^2{\rm v}_\bk }{d\tau^2} +\left(|w| k^2 - \frac{(\nu^2
-1/4)}{\tau^2} \right) {\rm v}_\bk = 0, \ee where \be \nu \simeq
\frac{3}{2} + \epsilon_1 + \frac{1}{2} \epsilon_2.  \ee The
appropriately normalized solution with the correct asymptotic
behavior is \be {\rm v}_\bk \to \frac{1}{\sqrt{2k_w}}e^{i k_w
\tau}, \ \ \ \ k_w/aH \to \infty, \ee where we have defined $k_{w}
\equiv \sqrt{-w}k$. Hence, up to a phase,
\be {\rm v}_\bk = \frac{\sqrt{\pi}}{2 } (-\tau)^{1/2} H_\nu^{(1)}
(-k_{w} \tau), \ee where $H_{\nu}^{(1)}$ is the Hankel function of
the first kind of rank $\nu$.
Since we are interested in the power spectrum of
this solution in the limit where all the modes are well outside
the horizon, only the dominant contribution of the asymptotic form
of the Hankel functions for $k_w/aH \to 0$ is considered, yielding
\be {\cal P}^{1/2}_{\cal R}(k) \equiv \sqrt{\frac{k^3}{2 \pi^2}}
|{\cal R}_{\bf k}| = 2^{ \nu} \frac{\Gamma(\nu)}{\Gamma(3/2)}
(-w)^{\frac{1-\nu}{2}} (1+\epsilon_1)^{1/2 - \nu} \left. \frac{
H}{8 \pi M_{\rm Pl}\sqrt{\epsilon_1}} \right|_{k=aH}.  \ee This is
the asymptotic value of the power spectrum of perturbations for
$k_w/aH \to 0$, written in terms of quantities evaluated at
horizon crossing $k=aH$. It corresponds to the result found for
$k$-inflation with $c_S^2 = -w $ (see the Appendix, Sec.~B). On
expanding in the slow-roll parameters it leads to
\bea
{\cal P}^{1/2}_{\cal R}(k) &=& (-w)^{-1/4} \left[1 -  (C + 1)
\epsilon_1 -\frac{1}{2} C \epsilon_2 \right] \left. \frac{ H }{2
\sqrt{2} \pi M_{\rm Pl}\sqrt{\epsilon_1}} \right|_{k=aH} \nn
\\
&=&  \left[1 -  ( C + 1 - \alpha) \epsilon_1 - \frac{1}{2} C \epsilon_2
\right]  \left. \frac{ H }{2 \sqrt{2} \pi M_{\rm Pl}
\sqrt{\epsilon_1}} \right|_{k=aH}, \ \ \ \
\alpha = \frac{1}{6}, \label{eq:scalarspec}
\eea
where $C\equiv-2 + \ln 2 + \gamma \simeq -0.72$ is a numerical
constant, $\gamma$ being the Euler constant originating in the
expansion of the gamma function. The parameter $\alpha$ vanishes
in SSFI.  Indeed, Eq.~(\ref{eq:scalarspec}) gives the same
spectrum as that of SSFI except for the factor $(-w)^{-1/4}\simeq
(1 + \alpha \epsilon_1)$. This factor will lead to a modification
of one of the second order consistency equations.

The spectrum of gravity waves in tachyon inflation is exactly as
in SSFI since in absence of anisotropic stress
gravity waves are decoupled from matter. The first
order result is (see e.g., \cite{Liddleetal}),
\be
{\cal P}^{1/2}_g(k) = [1 - (1+C)\epsilon_1] \frac{\sqrt{2}}{\pi}
\left. \frac{  H}{M_{\rm Pl} } \right|_{k=aH}.
\label{eq:tensorspec}
\ee

\subsection{Consistency relations}
\label{sec:prediction} \label{sub:F}

We now derive the consistency relations \cite{Liddleetal} linking
the tensor-scalar ratio $r$, the scalar spectral index $n$,
and the tensor spectral
index $n_T$ defined by \cite{Peiris}, \ba r &\equiv& \frac{{\cal
P}_g}{{\cal P}_{\cal R}} \label{eq:rdef},
\\
n &\equiv& 1+ \frac{d \ln {\cal P}_{\cal R}(k)}{d \ln k}, \\
n_T &\equiv& \frac{d \ln {\cal P}_g(k)}{d \ln k}.
\eea
The consistency equations are conditions on the observable parameters,
and they are thought to be distinctive of SSFI.
By finding deviations from the standard consistency
conditions we could in principle  find a way of distinguishing
tachyon inflation from SSFI.

As discussed above, only at next to lowest order does tachyon inflation
lead to deviations from SSFI. Indeed, as expected, to lowest order in the
slow-roll parameters $r$, $n$, and $n_T$ are identical to those of SSFI,
\ba
r &=& 16 \epsilon_1, \label{eq:r}
\\
n&=&1- 2 \epsilon_1 - \epsilon_2, \label{eq:n}\\
n_T&=&-2 \epsilon_1,
\eea
so the lowest order consistency relation is also the same,
\be
r= -8 n_T. \label{eq:c1}
\ee

Starting from Eqs.~(\ref{eq:scalarspec}) and
(\ref{eq:tensorspec}), we can compute $n$ and $n_T$ up to second
order in the slow-roll parameters,
\bea
n-1&=& -2\epsilon_1  -  \epsilon_2 - \left[2 \epsilon_1^2 +
(2 C +3 -2 \alpha ) \epsilon_1 \epsilon_2 + C \epsilon_2 \epsilon_3 \right], \\
n_T &=& -2 \epsilon_1 \left[1 +\epsilon_1 + (1+C)\epsilon_2 \right].
\label{eq:n_T}
\eea
This expression for $n_T$ is of course identical to that of SSFI,
whilst $n$ is different since $\alpha$ is nonzero, $\alpha=1/6$.
Also, the higher to lowest order expression for $r$ (see
Eq.~(\ref{eq:rdef})) is given by
\be
r = 16 \epsilon_1 \left[1 + C \epsilon_2 - 2 \alpha
\epsilon_1 \right].
\ee
The term $- 2 \alpha \epsilon_1 $ is absent in SSFI and
distinctive of tachyon inflation. By combining this expression
with Eq.~(\ref{eq:n_T}) and the lowest order results
Eqs.~(\ref{eq:r}) and (\ref{eq:n}) we get
\be
n_T =-\frac{1}{8} r \left[1 - \frac{1 - 2 \alpha}{16}
r + (1-n) \right]. \label{eq:c2}
\ee
This consistency relation is the next order version of
Eq.~(\ref{eq:c1}), and there is a clear deviation from SSFI which
is represented by a nonvanishing $\alpha$.  Note that this
deviation does not depend on any free parameter.  Similar
modifications of the consistency relations have been discussed in
$k$-inflation models \cite{GarriMuka} where the authors
generically find, at lowest order, $ r=-8 c_S n_T$, which
corresponds to our result with $c_S = (-w)^{1/2} \simeq 1$. Here
we have derived a next order consistency equation for the tachyon,
Eq.~(\ref{eq:c2}), which is different from the one of SSFI.
However, much larger departures can be caused by other mechanisms
\cite{Liddleetal,Knox,departures}.

\begin{figure}
\begin{center}
\includegraphics*[width=10cm]{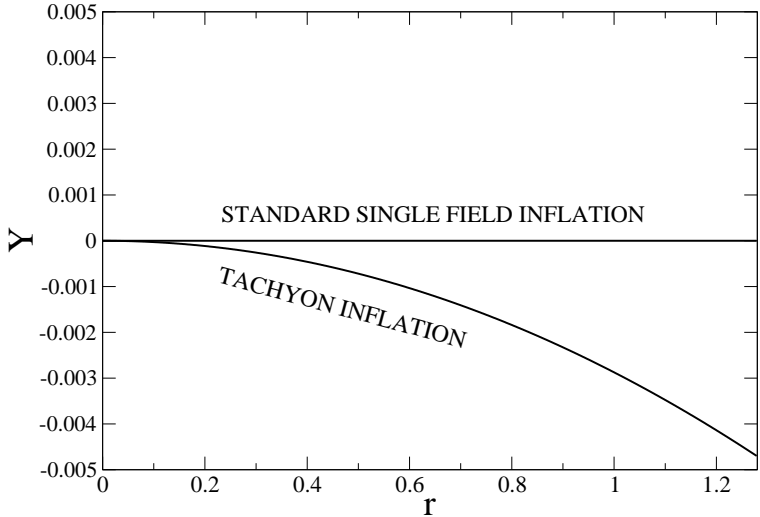}
\caption{The consistency relation
$Y =  - \frac{\alpha}{64}r^2 $,
for SSFI ($\alpha =0$) and tachyon inflation ($\alpha =1/6$).
In order
to be tested, we need $n$, $r$, and $n_T$, with a precision of
$\sim  10^{-3} r^2$.} \label{fig:cons}
\end{center}
\end{figure}
Figure~\ref{fig:cons} shows
\be
Y \equiv n_T +\frac{1}{8} r \left[1 - \frac{1}{16} r + (1-n)
\right] = - \frac{\alpha}{64}r^2
\ee
as a function of $r$. The equality $Y=0$ is satisfied in SSFI and
the deviation with $\alpha=1/6$ shown is a signature of tachyon
inflation.  However, in order to see the deviation predicted by
Eq.~(\ref{eq:c2}), $n$, $r$, and $n_T$ should be known to a
precision of $\sim 10^{-3}r^2$.  In \cite{Knox} the error on $n_T$
for future Cosmic Microwave Background observations is estimated.
Even for the largest possible values\footnote{As an upper limit on
the tensor-scalar ratio we take the $95 \%$ upper limit $r < 1.28$
from WMAP with no prior on the spectral index and its running, as
quoted in Ref.~\cite{Peiris}.} of $r$, $r \sim 1 $, this is too
large for the deviations predicted by the tachyon to be
observable, so that Eq.~(\ref{eq:c2}) will be very difficult to
test with planned experiments.

For completeness, we give expressions for the running of the
spectral indexes, $d n /d \ln k$ and $d n_T /d \ln k$, which are
the same as in SSFI,
\bea
\frac{d n}{d \ln k} &=& -2 \epsilon_1 \epsilon_2
- \epsilon_2 \epsilon_3, \\
\frac{d n_T}{d \ln k} &=&-2 \epsilon_1 \epsilon_2. \label{eq:dndk}
\eea
From Eq.~(\ref{eq:dndk}) one can derive a third consistency
equation, still identical to the SSFI one,
\be
\frac{d n_T}{d \ln k} = \frac{1}{8} r \left[ \frac{1}{8} r + (n-1) \right].
\ee

\section{Models of tachyon inflation}\label{sec:models}

We now study tachyon inflation for different potentials $V(T)$ and
extract $n$, $n_T$, $r$, and $dn/d \ln k$. We follow the standard
procedure \cite{RiottoLyth}: 1) for a given potential compute
$\epsilon_{1}$, $\epsilon_{2}$, and $\epsilon_2 \epsilon_{3}$ as a
function of $T$; 2) estimate $T_e$, the value of $T$ at the end of
inflation when $\epsilon_1(T_e)=1$; 3) calculate the number of
$e$-foldings as a function of the field $T$; 4) from
$\epsilon_{1}$, $\epsilon_{2}$, and $\epsilon_{3}$ calculate the
observable parameters as a function of $T$ and evaluate them at
$T(N_*)$. We consider only $T>0$.

\begin{figure}
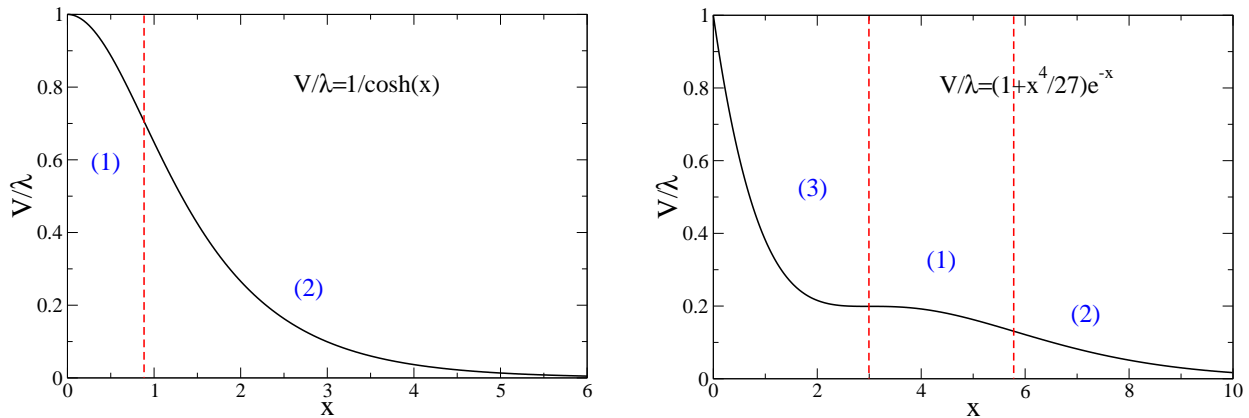

\centerline{ \scalebox{0.32}{
\includegraphics{coshplot.eps}}
\hspace*{0.5cm}
 \scalebox{0.32}{ \includegraphics{powplot.eps}}}
\caption{Illustrations of the regimes 1, 2, and 3, for two potentials.}
\label{fig:pot}
\end{figure}

As in SSFI (see e.g.~\cite{LiddleLyth}), we first introduce a
general classification scheme for models of tachyon inflation
where $V(T)$ satisfies conditions (\ref{eq:propsV}).  In order to
classify the available models we distinguish between three
possible regimes in which inflation may take place:
\begin{enumerate}
 \item $V'' \leq 0$, so that $6 \epsilon_1 \leq \epsilon_2$,
\item $0 < V'' < V'{}^2/ V$, so that $2 \epsilon_1 < \epsilon_2 <
6 \epsilon_1 $, \item $V'{}^2/ V \leq V''  $, so that $\epsilon_2
\leq 2 \epsilon_1 $.
\end{enumerate}
Fig.~\ref{fig:pot} shows these regimes for two potentials studied
below.
For several aspects and predictions, regime 1 is similar to small
field SSFI (although there $V'<0$ corresponds to $4 \epsilon_1
\leq \epsilon_2$). Regime 2 has similarities with large or chaotic
SSFI (when $0 < \epsilon_2 < 4 \epsilon_1$ in SSFI). Only in
regime 3 may $n$ be blue, though we shall see that the ``hybrid
inflation regime'' of SSFI ($n>1$ and large $n_T$) cannot be
probed.

Note that the parameter $X_0$ defined in Eq.~(\ref{eq:X0}) plays a
crucial r\^ole in the classification scheme introduced above, and
determines for which values of $x$ inflation occurs. Let $f(x)
\equiv V/\lambda$ so that
Eqs.~(\ref{eq:epsilonV})--(\ref{eq:etaV}) are just
\bea
\epsilon_1 &=& \frac{1 }{2 X_0^2} \frac{f'^2}{f^3}, \\
\epsilon_2 &=& \frac{1 }{X_0^2} \left(-2 \frac{f''}{f^2} +3
\frac{f'^2}{f^3} \right). \label{eq:f} \eea For the potentials we
consider, $x_e\equiv T_e/T_0$ increases with $X_0$ which itself
must be sufficiently large that the slow-roll conditions are
satisfied, $\epsilon_1, \epsilon_2 \ll 1$, in some range of $x$.
Finally, from Eq.~(\ref{eq:NNN}) the number of $e$-foldings is \be
N_* = X_0^2 \int_{x_*}^{x_e} \frac{f^2}{ |f'|} dx,
\label{eq:Nagain} \ee which is a monotonically decreasing function
of $x_*$. Thus for a given $N_*$ (i.e., $N_* \simeq 60$),
increasing $X_0$ also implies increasing of $x_*$.

Before discussing different examples, note that from the recent
limits on the amplitude of scalar and tensor perturbations
\cite{Peiris}, it is possible to bound $\lambda$. At lowest order
in $\epsilon_1$,
\be
{\cal P}_{\cal R} \simeq \frac{V_*}{24 \pi^2 M_{\rm Pl}^4
\epsilon_{1*}} = \frac{2 V_*}{3 \pi^2 M_{\rm Pl}^4 r} \simeq
2.95 \times 10^{-9} \times (0.71 + 3 \times 0.01) , \label{eq:COBE}
\ee
where $0.71 + 3 \times 0.01$ is the $3 \sigma$ upper limit on the
amplitude (at $k=0.002 \ {\rm Mpc}^{-1}$). We have
\be
V_*^{1/4} \simeq 1.83 \times (r/0.1)^{1/4} \times 10^{16} {\rm
GeV},
\ee
and from the upper bound on gravitational waves $ r < 1.28$,
\be
\lambda^{1/4} \lsim 3.46 \times 10^{16} {\rm GeV}. \label{eq:constlambda}
\ee
For $X_0 \gg1$ this implies
\be
T_0 \gg 2.5 \times 10^4 \ell_{\rm Pl} \equiv 2.5 \times 10^4 \sqrt{G}.
\ee

\subsection{Inverse cosh potential}

A number of different tachyon potentials (with the properties mentioned
in the introduction) have been derived in the string
theory literature.  Initially we consider the potential \cite{action,bsft}
\be
V = \frac{\lambda}{\cosh x},
\ee
which was studied in a
cosmological context in \cite{cosmocosh}.
At first order in the slow-roll approximation, which is valid for
$X_0 \gsim 3$,
\bea \epsilon_1 &=& \frac{1 }{2
X_0^2} \frac{\sinh^2 x}{\cosh x}, \\
\epsilon_2 &=& \frac{1
}{X_0^2} \frac{\cosh^2 x + 1 }{\cosh x}, \\
\epsilon_2
\epsilon_3 &=&  \frac{1}{X_0^4} \frac{\sinh^4 x}{\cosh^2 x}.
\label{eq:slowcosh}
\eea
For $x \lsim 1$, inflation takes place in the regime 1 of our
classification, while for $x \gsim 1$, it takes place in the
regime 2 (see Fig.~\ref{fig:pot}). Inflation ends in regime 2 when
$x_e \simeq \ln(4X_0^2) \gsim \; 4$.
The spectrum of scalar
and tensor perturbations are both red, while the running of the
scalar spectral index is always negative,
\bea
n-1&=& - \frac{2}{X_0^2} \cosh x,
\\
\frac{d n}{d \ln k}&=& - \frac{2}{X_0^4} \sinh^2x , \\
 r = - 8
n_T&=& \frac{8}{X_0^2} \frac{\sinh^2 x}{\cosh x}.
\eea

To compare the predictions for the inverse cosh
potential with data, $x_*$ must first be determined. From
Eq.~(\ref{eq:Nagain})
\be
N_* = X_0^2 \int_{x_*}^{x_e} \frac{1}{\sinh x} dx  = X_0^2 \left[
\ln \left( \frac{e^{x_*}+1}{e^{x_*}-1} \right) - \ln \left(
\frac{e^{x_e}+1}{e^{x_e}-1}  \right) \right], \label{eq:nstar}
\ee
so that
\be
\tanh (x_*/2) = \exp(-N_*/X_0^2) \tanh(x_e/2) \simeq \exp(-N_*/X_0^2).
\label{eq:Nofxstar}
\ee
Thus $X_0^2 \sim N_*$ distinguishes between regime 1 and 2. From
the last equality of Eq.~(\ref{eq:Nofxstar}) one finds
$$
\sinh x_* = \left[ \sinh (N_*/X_0^2) \right]^{-1}, \ \ \ \ \cosh x_*
=  \left[\tanh (N_*/X_0^2)  \right]^{-1},
$$
so that the observables $n$, $dn/d \ln k$ and $r$ can be reexpressed
in terms of $N_*$,
\bea n-1  &=& -\frac{2}{X_0^2}
\frac{1}{\tanh(N_*/X_0^2) }, \label{eq:spectralcosh}
\\
\frac{d n }{d \ln k} &=& - \frac{2}{X_0^4} \frac{1}{\sinh^2
(N_*/X_0^2)}, \label{eq:runningcosh}
\\
r =-8 n_T &=& \frac{16}{X_0^2} \frac{1}{\sinh (2N_*/X_0^2)}
\label{eq:rcosh}.
\eea
The running of the spectral index is very small in
this model: at the $2\sigma$ level
\be
0 > \frac{d n}{d \ln k}= -(1-n) \frac{r}{8} \gsim - 0.002.
\ee
The potential $V=\lambda (1+x)e^{-x}$ that has been discussed in
the string theory literature \cite{open} has very similar
properties to this inverse cosh potential.

\subsection{Exponential potential}
\label{subsec:exp}

Another potential which has been derived and studied a great deal
in the string and tachyon literature (see for instance
\cite{Sami,Sen2}) is
\be
V(x) = \lambda \exp(-x).
\ee
From Eq.~(\ref{eq:NNN}) we find
\be
N_* = X_0^2 (e^{-x_*} -
e^{-x_e}),
\label{eq:Nexp}
\ee
and to first order in the slow-roll
approximation
 \be
 \epsilon_1 = \frac{\epsilon_2}{2} = \frac{1 }{2
X_0^2} e^{x} \; , \qquad \epsilon_2 \epsilon_3 =
\frac{e^{2x}}{X_0^4},
\label{eq:slowexp}
\ee
so that this potential is on the border of regimes 2 and 3.
Inflation ends when $e^{x_e}=2 X_0^2$ and since $x_*\geq 0$ it
follows from Eq.~(\ref{eq:Nexp}) that $X_0^2 \geq N_* + 1/2 \gsim
N_*$.  This should be contrasted with the case of the inverse cosh
potential: there, a sufficient number of $e$-foldings can be
obtained for any value of $X_0$.

The
spectrum of scalar and tensor perturbations are both red,
\bea
n-1&=& - \frac{2}{(N_*+1/2)}, \label{eq:nexp}
\\
\frac{d n}{d \ln k} &=& - \frac{2}{(N_*+1/2)^2}, \\
r = -8 n_T &=& \frac{8 }{(N_*+1/2)},\label{eq:rexp}
\eea
and the running of the spectral index is negative and very small,
\be
0 > \frac{d n}{d \ln k} = -\frac{1}{2} (n-1)^2 = - (1-n) \frac{r}{8}
\gsim -0.002
\ee
at the $2\sigma$ level. One should not be surprised that these
results are the same as those of chaotic SSFI with
$W(\phi)=\frac{1}{2} m^2 \phi^2$:  when using the change of
variables Eq.~(\ref{eq:change1}) for the linearized action this
potential indeed corresponds to $V=\lambda \exp(-x)$.  Note also
that in the limit $x \to \infty$ the inverse $\cosh$ potential
reduces to this exponential potential: as
expected, Eqs.~(\ref{eq:spectralcosh})--(\ref{eq:rcosh}) indeed
coincide with Eqs.~(\ref{eq:nexp})--(\ref{eq:rexp}) in the $X_0
\rightarrow \infty$ limit.
\begin{figure}
\begin{center}
\includegraphics*[height=8cm]{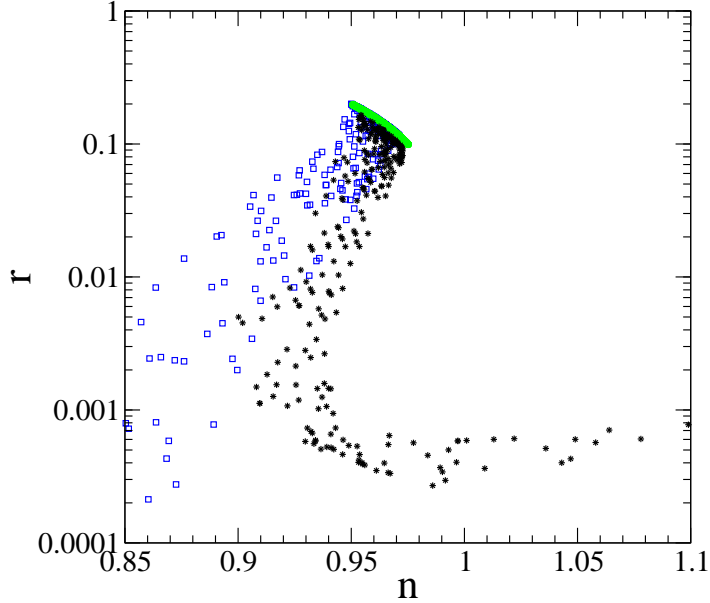}
\caption{Comparison of the predictions of the inverse cosh
potential (squares, blue), the potentials $V = \lambda
e^{-x}$ (circle, green), and $V = \lambda (1+3^{-3} x^4) e^{-x}$
(stars, black). }
\label{fig:log}
\end{center}
\end{figure}
Fig.~\ref{fig:log} shows a random sampling of the $(n,r)$-plane
predictions for the inverse cosh potential and $V=\lambda
\exp(-x)$.  We have varied both $X_0$ and the number of
$e$-foldings $N_*$ with $40 \le N_* \le 70$.

\subsection{Inverse power-law potential}
\label{subsec:power}

In the remainder of this section we no longer work with tachyon potentials
derived in string theory.  Instead we introduce two
phenomenological potentials which still satisfy the properties
mentioned in the introduction. The reason for doing this
is that they probe region 3 of our
classification.

The first potential is an inverse power-law
potential such as
\be
V=\frac{\lambda}{ (1+x^4)}.
\ee
Note that a similar inverse power-law potential, $V \propto 1/x^\alpha$,
with $\alpha >0$, has been considered in \cite{SolM,Finelli}.
At first order in the slow-roll approximation, which is valid for
$X_0 \gsim 7 $,
\bea
\epsilon_1 &=& \frac{8x^6}{X_0^2(1+x^4)}, \\
\epsilon_2 &=& \frac{8x^2(x^4+3)}{X_0^2(1+x^4)}, \\
\epsilon_2 \epsilon_3 &=& \frac{64 x^4(x^8+3)}{X_0^4(1+x^4)^2}.
\eea
As opposed to all the potential studied above, when $x$ is large
($x>3^{1/4}$) inflation occurs in regime 3 of our classification
where we expect an important contribution of gravity waves.
However, the scalar spectral index is red since $-\epsilon_2 < 2
\epsilon_1$ for all $x$.  To leading order in $1/X_0$, inflation
ends when
\be
x_e^2 = \frac{X_0^2}{8} + {\cal O}(X_0^{-2}),
\ee
and from Eq.~(\ref{eq:Nagain}) the number of $e$-foldings is
\be
N_* =  \frac{X_0^2}{8} \left( \frac{1}{x_*^2}-  \frac{1}{x_e^2}
\right),
\ee
so that
\be
x_*^2 = \frac{X_0^2}{8(1+N_*)} [1+ {\cal O}(X_0^{-2})].
\ee

The spectral indices are red and the running is negative,
\bea
n-1 &=& - \frac{24 x_*^2}{X_0^2} = -\frac{3}{1+N_*} [1 + {\cal
O}(X_0^{-2})], \label{eq:pown}\\
\frac{dn}{d\ln k} &=& -\frac{192 x_*^4}{X_0^4} = -
\frac{3}{(1+N_*)^2} [1+ {\cal O}(X_0^{-2})], \\
r = - 8 n_{T} &=& \frac{128 x_*^6}{X_0^2(1+x^4)} =
\frac{16}{1+N_*}  [1+{\cal O}(X_0^{-2})]. \label{eq:powr}
\eea
As for the potentials considered previously, the running of $n$ is very small
\be \frac{d n}{d \ln k} = -\frac{1}{3} (n-1)^2 \gsim -0.002,
\ee
at the $2\sigma$ level.
\begin{figure}
\begin{center}
\includegraphics*[height=8cm]{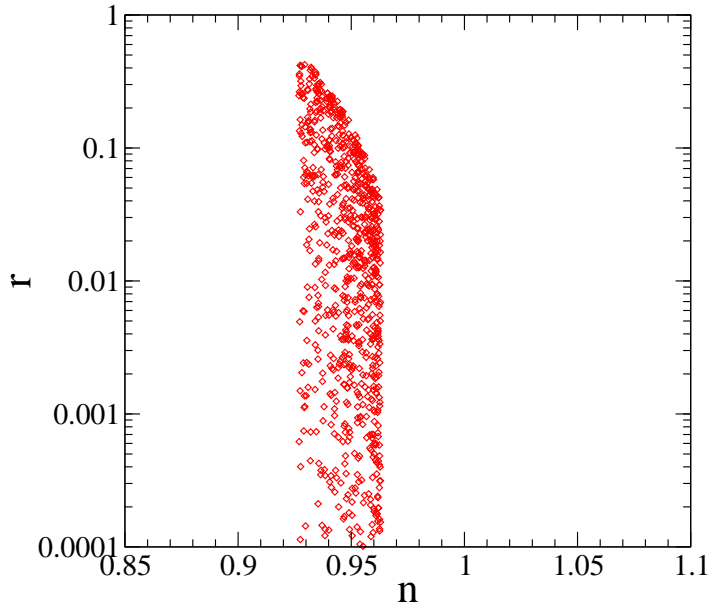}
\caption{The $(n,r)$-plane for the inverse power-law potential. }
\label{fig:log2}
\end{center}
\end{figure}
Fig.~\ref{fig:log2} plots a random sampling of the $(n,r)$-plane
for $40 \le N_* \le 70$ showing the more important contribution of
gravity waves relative to the inverse cosh and exponential
potentials.

\subsection{Potentials giving $n \geq 1$}

All the potentials considered above give $n<1$.  It is legitimate
to ask if potentials satisfying the conditions (\ref{eq:propsV})
can give $n \ge 1$. This requires
\be
-\epsilon_2 \geq 2 \epsilon_1 ,
\label{eq:condg}
\ee
and the equality is safistifed for $V=\lambda/(1+x)$.  More generally
the power-law potentials
\be
V = \frac{\lambda }{1+x^p}
\label{eq:Vpow}
\ee
with $p \leq 1$ give $n \geq 1$.  However,  in the absence of
other physics, inflation never ends for these potentials since
$\epsilon_1$ is a decreasing function of $x$. In the class of
exponential potentials
\be
V = \lambda e^{-x^p},
\label{eq:Vp}
\ee
the condition Eq.~(\ref{eq:condg}) requires $p\leq 1$, and $n\geq
1$ when $x_* \leq x_{c}=[(1-p)/p]^{1/p}$.   Once again, other
physics such as the coupling to another field is required to end
inflation, similarly to hybrid models.

If we remain with models in which inflation ends naturally we can
modify $V$ to obtain $n \geq 1$.  An example is the potential
\be
V = \lambda \left( 1 + (p-1)^{(1-p)} x^p \right) e^{-x},
\label{Vcra}
\ee
which is not string theory inspired.  For $p=1$ the predictions
are very similar to those of the inverse cosh potential. For $p>1$
the potential explores regime 3 of our classification (see
Fig.~\ref{fig:pot}). By construction, when $x=p-1$, $V'=V''=0$ so
that $\epsilon_i=0$ leading to $n=1$ and $r=0$. However, for $x_*
\leq p-1$ we are in regime 3, $n \geq 1$ and $r \propto
\epsilon_1$ is very small. Hence, as is confirmed in
Fig.~\ref{fig:log}, this potential leads to a blue scalar spectral
index with very small $r$. This regime is probed by small $X_0$.
For large $X_0$ the predictions are the same as the inverse cosh
potential (again see Fig.~\ref{fig:log}).

\section{Discussion and conclusion}

\begin{figure}
\begin{center}
\includegraphics*[height=13cm]{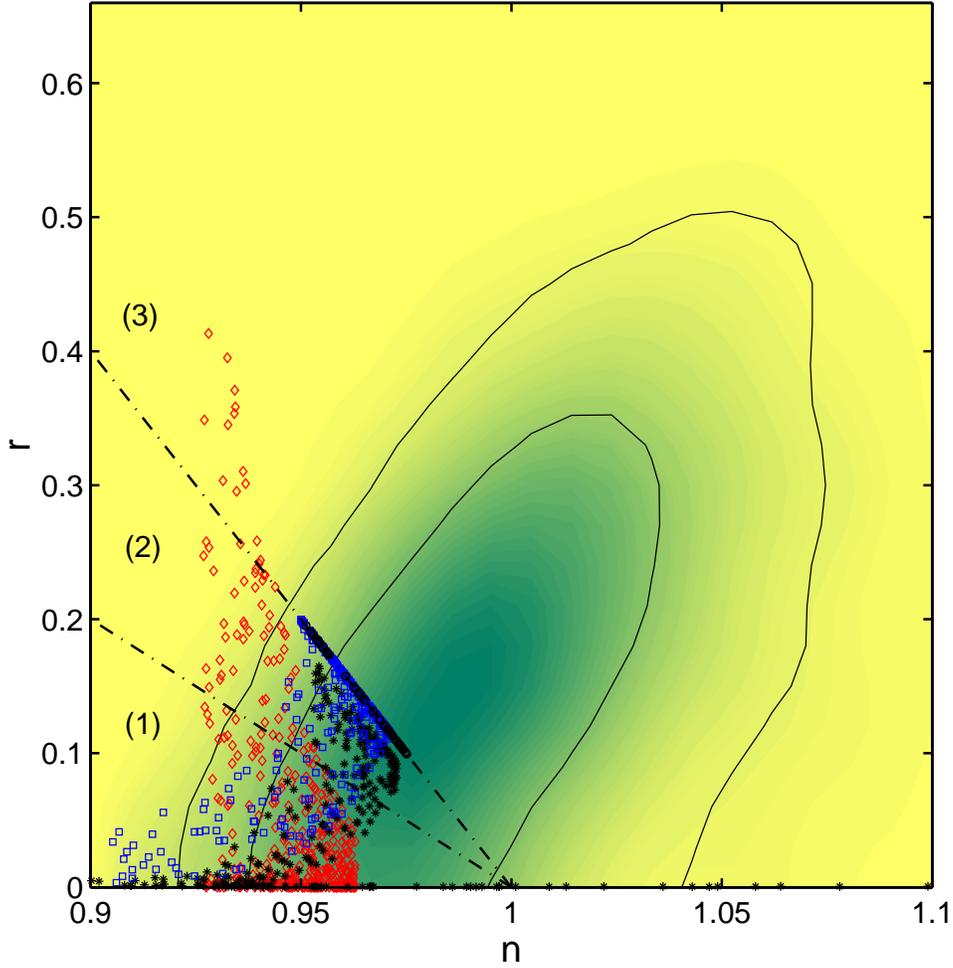}
\caption{Models of tachyon inflation compared to the 2-dimensional
likelihood contours (at $1\sigma$ and $2\sigma$) on the
$(n,r)$-plane.  The points represent the result of a random
sampling from three different tachyonic potentials: inverse
$\cosh$ potential (squares, blue), $V=\lambda e^{-x}$ (circles,
black), the potential $V=\lambda(1+x^4/27)e^{-x}$ (stars, black),
and inverse power-law potential (diamonds, red). The two dashed
lines correspond to the limits between the three different regimes
of inflation. The likelihood contour comes from the analysis of
S.~Leach and A.~Liddle \cite{Sam}. }
\label{fig:wmap}
\label{fig:sam}
\end{center}
\end{figure}
In the previous section we have described the inflationary
predictions of several tachyon potentials. They generally have a
red spectrum of scalar perturbations with a negative and very
small running of the scalar spectral index. For specific choices
of potential such as that given in Eq.~(\ref{Vcra}), blue spectra
can be obtained with very small $r$. It is interesting to compare
these predictions with the current data and to see whether it is
possible to discriminate between SSFI and tachyon inflation.

Since there is a very small running, it is legitimate to ask how
well motivated it is to introduce this new higher-order parameter
(and hence $\epsilon_3$) when comparing our models with data.
Motivated by the discussion of Leach and Liddle \cite{Sam}, we
neglect $\epsilon_3$ and use their likelihood analysis to
constrain the first two parameters $\epsilon_1$ and $\epsilon_2$.
We consider the same four potentials as in figures \ref{fig:log}
and \ref{fig:log2} and compare their $(n,r)$-plane predictions
with the 2-dimensional likelihood contours (at $1\sigma$ and
$2\sigma$).  Results are shown in Fig.~\ref{fig:sam}.

For the inverse cosh potential (squares, blue) inflation can take
place in both regimes 1 and 2. For a large set of parameters $N_*$
and $X_0$ (excluding very small $X_0$), the predictions are well
inside the $2\sigma$ contour. There are non negligible gravity
waves for large $X_0$, though for the range of $N_*$ given above,
$r \lsim 0.2$. When $X_0 \to \infty$, the predictions concentrate
on the line $\epsilon_2 = 2 \epsilon_1$ which are just those of
the exponential potential $V=\lambda e^{-x}$ (circles, black). The
inverse power-law potential (diamonds, red) can occupy regimes 1,
2, and 3, and leads to a large contribution of gravity waves,
although $r \lsim 0.2$ in the region not excluded by current data.
The potential $V=\lambda (1+x^{4}/27 ) e^{-x}$ (stars, black)
occupies much of the region of the inverse cosh potential as well
as yielding blue spectra for negligible $r$.

All the presented models seem to be consistent with the data.
Hence, the first-year WMAP results
are still too crude to constraint significantly the region of
parameters. On the other hand, we still lack of information about
the mechanism of reheating that could take place after tachyon
inflation, leaving us with a large uncertainty on $N_*$. Progress
can be made by better estimating this particular parameter.

Our results point to the fact that it is difficult to distinguish
between a model where inflation is driven by a Klein-Gordon scalar
field or by some other field satisfying a non standard action.
However, none of the potentials we have considered in our
analysis, which are those where inflation ends naturally, lead to
both a blue scalar spectral index and large gravity wave spectrum.
Therefore for these potentials a large region in the $(n,r)$-plane
is not probed by tachyon inflation. This corresponds, in SSFI, to
the region occupied by hybrid inflation. Detection of $n>1$ and
large $r$, or of a large running of $n$, can lead to the exclusion
of tachyonic inflation.

In this paper we have discussed tachyon inflation using a
phenomenological approach. We have presented a new exact solution
of the tachyon-gravity equations which smoothly interpolates
between the inflationary and the dust-like regime, and have shown
that one of the consistency relations differs from that of
standard inflation. It will be difficult to use this equation to
discriminate between standard inflation and tachyon inflation with
planned observations, but things may improve in the future.
However, this modified consistency relation may be useful to
constraint other $k$-type inflationary models as we have discussed
in the Appendix. Finally we compared the predictions of tachyon
inflation with current data. None of the models presented here can
be excluded by the data. We conclude that if the future data point
towards small and chaotic single field inflation, it may be
difficult to discriminate between tachyon inflation and standard
single field inflation, unless other distinguishing criteria
appear.

\section*{Acknowledgments}

It is a pleasure to thank R.~Durrer, S.~Leach, J.~Martin, and
R.~Rivers for useful discussions and comments, and T.~Evans for
help with Maple. We acknowledge S.~Leach and H.~Peiris for
providing us with their Markov Chain analyses.

\appendix

\section{Tachyon perturbations}\label{sec:pert}

In this appendix we recall two important properties of a tachyon
dominated universe. First, as for a single standard scalar field
dominated universe, the perturbations produced during tachyon
inflation are adiabatic. The entropy perturbation is present but
it is negligible on large scales, $k/(aH) \to 0$. This ensures
that the large scale perturbations produced during a tachyon
inflationary phase are adiabatic, as in SSFI. However, the analogy
with the Klein-Gordon scalar field stops here:
in contrast to the scalar field, the tachyon can cluster on small
scales when $\dot{T}^2 \to 1$, as was shown in
\cite{FKS,Wasserman} (see also \cite{GarriMuka} for the same
derivation in the context of $k$-inflation). Here we rederive this
result making use of a gauge invariant formalism, and show that
the result relies on the fact that the tachyon and the scalar
field have different entropy perturbations on small scales.

We study perturbations about the homogeneous solution $T(t)$,
considering the coupled system of tachyon and metric
perturbations. The perturbed metric is defined by
\be
ds^2 =  -(1+2A)dt^2 + 2 a \partial_i B dx^i dt + a^2  [(1-2 \psi)
\delta_{ij} +2 \partial_i \partial_j E ] dx^i dx^j.
\ee
As in the case of a scalar field, the tachyon has no anisotropic
stress and perturbations can be described by two independent
variables (e.g. $\delta T$ and $\psi$ in longitudinal gauge).  We
now derive evolution equations of the density contrast.

It is convenient to treat the tachyon as a perturbed non adiabatic fluid.
%
Thus, we can define the intrinsic entropy
perturbation as \cite{SasakiKodama}
\be
\Gamma \equiv \frac{\delta P}{P} - \frac{c_A^2}{w} \frac{\delta
\rho}{\rho} .
\ee
For the tachyon
\be
P \Gamma = \frac{2}{3} \frac{(\ln V)'}{H \dot T} \rho
\Delta = (c_S^2  - c_A^2)\rho  \Delta , \label{eq:entropytachyon}
\label{eq:entropytdu}
\ee
where, using the notation of \cite{SasakiKodama}, $\Delta $ is the
gauge invariant density perturbation on the comoving hypersurface
relative to the tachyon fluid,
\be
\Delta \equiv \frac{\delta \rho}{\rho}  +3H \dot T \delta T .
\ee
The last equality of Eq.~(\ref{eq:entropytachyon}) is
valid, in general, for any $k$-field \cite{GarriMuka}.

In the fluid formalism, we find a second order differential
equation for $\Delta$ \cite{SasakiKodama},
\be
\frac{d^2 \Delta}{d\tau^2} - \HH [3(2 w - c_A^2) -1] \frac{d
\Delta}{d \tau} + 3 \HH^2 \left(\frac{3}{2} w^2 - 4 w -
\frac{1}{2} + 3 c_A^2 \right) \Delta + c_A^2 k^2 \Delta = - w  k^2
\Gamma, \label{eq:Deltaevol2}
\ee
where $\tau$ is conformal time, $d\tau \equiv dt/a$, and $\HH
\equiv  (da/d \tau)/a$. By using Eq.~(\ref{eq:entropytachyon}) we
obtain the evolution equation for the density contrast of a
tachyon fluid,
\be
\frac{d^2 \Delta}{d\tau^2} - \HH [3(2 w - c_A^2) -1] \frac{d
\Delta}{d \tau} + 3 \HH^2 \left(\frac{3}{2} w^2 - 4 w -
\frac{1}{2} + 3 c_A^2 \right) \Delta + c_S^2 k^2 \Delta = 0.
\label{eq:Deltaevol}
\ee
Here the sound horizon $c_S/(aH)$, from the Laplacian term of
Eq.~(\ref{eq:Deltaevol}), should be contrasted with
$c_A/(aH)$  in the case of an adiabatic fluid where
$\Gamma=0$. The difference is due to the non-zero entropy
perturbation $\Gamma$ which, in the fluid formalism of
Eq.~(\ref{eq:Deltaevol2}), acts as a source for the evolution
equation for the energy density perturbation.

For a standard scalar field $c_S^2 = 1$ while for the tachyon $c_S^2 = -w$.
On large scales, when $k/(aH) \to 0$, Eq.~(\ref{eq:Deltaevol2}) is the same
for the two fields.
In particular, as for a standard scalar field,
in a tachyon dominated universe curvature
perturbations are conserved and are not affected by entropy
perturbations. The curvature perturbation on the comoving
hypersurfaces is defined by
\be
{\cal R} = \psi + H \frac{\delta T}{\dot T}. \label{eq:curv}
\ee
On large scales this is sourced only by the entropy perturbation according
to \cite{Gamma}
\be
\dot {\cal R} \simeq - 3 H^2 \frac{P}{\dot \rho} \Gamma.
\ee
However, combination of the perturbed energy and momentum
constraint equations gives
\be
\frac{k^2}{a^2} \Psi = - 4 \pi G \rho  \Delta ,
\ee
where $\Psi$ is the Bardeen potential $\Psi = \psi - \HH(B-a\dot
E)$. Using Eq.~(\ref{eq:entropytachyon}), the entropy in a tachyon
dominated universe becomes
\be
P \Gamma =  \frac{w + c_A^2}{4 \pi G } \frac{k^2}{a^2} \Psi \simeq
0, \ \ \ \ {\rm for } \ \ k/(aH) \to 0,
\ee
and the curvature perturbation ${\cal R}$ is conserved on large
scales.

However, different $c_S^2$'s imply
a different clustering behaviour on small scales, where the Laplacian term
becomes important.
It is well known that a standard
scalar field (such as quintessence) cannot cluster on length
scales smaller than $H^{-1}$, its Jeans length being of the order
$L_J \sim c_S/(aH) = (aH)^{-1}$. For tachyon matter, however, $c_S^2 =-w$
and the effective Jeans length for scalar perturbations is only of
order $L_J = (aH)^{-1}(1-\dot T^2)^{1/2}$, which can be very small.
Thus the tachyon can cluster on small scales when $\dot T \to 1$.

\section{Generalized action}

Here we consider the action
\be
S_T = - \int d^4 x \sqrt{-g}  V(T)  \left( 1 +
g^{\mu \nu} \partial_{\mu} T
\partial_{\nu} T  \right)^q, \label{eq:actionq}
\ee
with the aim of clarifying the origin of the differences between tachyon
inflation $q=1/2$ and SSFI $q=1$.

From Eq.~(\ref{eq:actionq}) it follows that
\ba
P &=& -V(1-\dot{T}^2)^q,
\\
\rho &=& V (1-\dot{T}^2)^{q-1} \left[ 1 - \dot{T}^2(1-2q) \right],
\label{eq:rhoq}
\\
w &=&  - \frac{(1-\dot{T}^2)}{1 - \dot{T}^2(1-2q) }, \ea and the
equation of motion for $T$ is
\be
(2q) \frac{\ddot{T}}{1-\dot{T}^2} + 3 (2q) H
\frac{\dot{T}}{1-\dot{T}^2(1-2q)} + ({\rm ln}V)'=0.
\ee
Inflation ends when
\be
\dot{T}^2 = \frac{1}{q+1}.
\ee

The evolution equation for density perturbations about
this homogeneous and isotropic
background is given by Eq.~(\ref{eq:Deltaevol}) where
\be
c_S^2
= 1 + (q-1) \frac{2 \dot{T}^2}{1 + \dot{T}^2(1-2q)}
= 1 + (q-1) \frac{2(1+w)}{1+(1-2q)(2w+1)}
.
\label{eq:Ceff}
\ee
For $q=1/2$, $c_S^2=-w$; and for $q=1$, $c_S^2=1$ as
discussed after Eq.~(\ref{eq:Deltaevol}).

It is straightforward to calculate
the number of $e$-foldings and the slow-roll parameters.
From the Friedmann equation with energy density $\rho$ given in
Eq.~(\ref{eq:rhoq}) it follows that
\be
dN = - \frac{3}{2} \frac{H^3}{H'}
\left[\frac{2q}{1 - \dot{T}^2(1-2q)} \right] dT,
\ee
so that by definition (see Eq.~(\ref{eq:defeps2})),
\be
\epsilon_1 \equiv  \frac{1}{H} \frac{dH}{dN} = \frac{3}{2}
\left[\frac{2q \dot{T}^2}{1 - \dot{T}^2(1-2q)} \right].
\label{eq:epsq}
\ee
Thus to leading order
\be
\epsilon_1 \simeq \frac{1}{2q} \frac{M_{\rm Pl}^2}{2} \frac{V'^2}{V^3}.
\label{eq:eps1q}
\ee

In the calculation of density perturbations, the pump field is now
\cite{GarriMuka}
\ba
z & \equiv &\frac{a(\rho+P)^{1/2}}{c_S H} =   \sqrt{3}a \sqrt{2q}
\dot{T} \left[
\frac{1+\dot{T}^2(1-2q)}{(1-\dot{T}^2)(1-\dot{T}^2(1-2q))}
\right]^{1/2} \label{eq:step1}
\\
&=&  a  \sqrt{2 \epsilon_1} \left({ 1 - \frac{2}{3}
\epsilon_1 } \right)^{-1/2}
 \left[ 1 + \frac{4}{3}
\frac{\epsilon_1}{2q}(1-2q) \right]^{1/2}. \label{eq:step2}
\ea
For $q=1$, Eqs.~(\ref{eq:step1}) and (\ref{eq:step2})
correctly reduce to
\be
z_{\rm SSFI} = \frac{\sqrt{6}a \dot{T}}{\sqrt{1+\dot{T}^2}} =
\frac{a\dot{\phi}}{M_{\rm Pl} H} = a \sqrt{2 \epsilon_1},
\ee
where the second equality is obtained using Eqs~(\ref{eq:change1})
and  (\ref{eq:change2}). Thus
\be
z_{\rm SSFI} = z
 \left({ 1 - \frac{2}{3}\epsilon_1 } \right)^{1/2} \left[ { 1 + \frac{4}{3}
\frac{\epsilon_1}{2q}(1-2q)} \right]^{-1/2},
\ee
which is the generalisation of Eq.~(\ref{eq:zcomp}). On
differentiation and using Eq.~(\ref{eq:epsq}) we find
\be
U = U_{\rm SSFI} + a^2 H^2  \epsilon_1 \epsilon_2
\left( \frac{1-q}{q} \right) +
{{\cal O}}(\epsilon_i^2),
\ee
again showing clearly the $q=1$ limit.

The expression for ${\cal P}_{\cal R}(k)$ is exactly as given in Eq.~(\ref{eq:scalarspec})
but with $w$ replaced by $-c_S^2$ which, to leading order, is given by
\be
c_S^2=1- \frac{2(q-1)}{3q} \epsilon_1 + {{\cal O}}(\epsilon_1^2).
\ee
A straightforward calculation then shows that the general expression for
$\alpha$ introduced in Eq.~(\ref{eq:scalarspec}) is
\be
\alpha = \frac{1-q}{6q},
\label{eq:alphaff}
\ee
so that $\alpha=0$ in SSFI and $\alpha=1/6$ in tachyon inflation.
Thus the consistency relations given in Section \ref{sub:F}
remain valid as a function
of $q$ provided the slow-roll expansion holds.
The consistency relation
\be
 n_T + \frac{r}{8} \left[1- \frac{r}{16}+(1-n) \right] \equiv
 Y = - \frac{\alpha r^2}{64}
\ee
can constrain $\alpha$, and hence $q$, for large $r$. For example,
when $r \simeq 1$ the error on $Y$ is predicted to be of the order
of $\sim 0.007$ \cite{Knox}. In that case values of $\alpha \gsim
1/2$ corresponding to $q \lsim 1/4 $ could be tested.  However, we
stress that only $q=1/2,1$ are physically relevant values here.

For a given potential $V(T)$, the inflationary predictions of
Section \ref{sec:models} depend on $q$. From Eq.~(\ref{eq:eps1q})
this $q$ dependence can simply be obtained by replacing
$\epsilon_i \rightarrow \epsilon_i=\epsilon_i/(2q)$ in the
calculations of that section.  (This factor of $1/(2q)$ originates
from action Eq.~(\ref{eq:actionq}):  when expanded to first order
in $\dot{T}^2$ its kinetic term differs by a factor of $1/(2q)$
from kinetic term the SSFI action, Eq.~(\ref{eq:actionSSFI}).)
 For example, consider the
exponential potential $V=\lambda \exp(-x)$ of section
\ref{subsec:exp}.  The number of $e$-foldings is now
\be
N_* = \sqrt{2q} X_0^2 (e^{-x_*} - e^{-x_e}), \label{eq:Nq}
\ee
and to first order in the slow-roll
approximation
\be
\epsilon_1 = \frac{\epsilon_2}{2} = \frac{1 }{ [(2q) X_0^2]}
\frac{e^{x}}{2} \; , \qquad \epsilon_2 \epsilon_3 =
\frac{e^{2x}}{[(2q) X_0^2]^2}. \label{eq:slowexpA}
\ee
Inflation ends when $e^{x_e}=2 (2q) X_0^2$ and Eq.~(\ref{eq:Nq})
imposes $(2q) X_0^2 \geq (\sqrt{2q}N_* + 1/2)$. The $q$-dependent
spectrum of scalar and tensor perturbations are given by
\bea
n-1&=& - \frac{2}{(\sqrt{2q} N_*+1/2)}, \label{eq:nexpbis}
\\
\frac{d n}{d \ln k} &=& - \frac{2}{(\sqrt{2q} N_*+1/2)^2}, \\
r = -8 n_T &=& \frac{8 }{(\sqrt{2q} N_*+1/2)}.\label{eq:rexpbis}
\eea
For this potential and for $N_* \gsim 40$, the
WMAP data constrains $q \gsim 0.16$, at the 2$\sigma$ level.

\end{document}